\documentclass[journal=jpcb,manuscript=article]{achemso}
\setkeys{acs}{articletitle = true}

\usepackage[version=3]{mhchem} % Formula subscripts using \ce{}
\usepackage[T1]{fontenc}       % Use modern font encodings

\usepackage[usenames, dvipsnames]{color}
\usepackage[table]{xcolor}
\usepackage{nameref}
\usepackage{graphicx}
\usepackage{subfigure}

\definecolor{Gray}{gray}{0.9}
\newcolumntype{g}{>{\columncolor{Gray}}c}
\newcolumntype{P}[1]{>{\centering\arraybackslash}p{#1}}

\usepackage{amsmath,amssymb}
\usepackage{siunitx}
\usepackage{hyperref}
\author{George Stan}
\affiliation[University of Cincinnati]
{Department of Chemistry, University of Cincinnati, Cincinnati, Ohio 45221, USA}
\email{george.stan@uc.edu}
\author{George H. Lorimer}
\email{glorimer@umd.edu}
\affiliation{Center for Biomolecular Structure and Organization, Department of Chemistry and Biochemistry,
University of Maryland, College Park, Maryland 20742, USA}
\author{D. Thirumalai}
\email{dave.thirumalai@gmail.com}
\affiliation[UT Austin]{Department of Chemistry and Department of Physics, University of Texas, Austin, Texas 78712, USA} 
\title[An \textsf{achemso} demo]
{Friends in need: how chaperonins recognize and remodel proteins that require folding assistance}

\begin{document}

\newpage
\baselineskip 22pt

%%%%%%%%%%%%%%%%%%%%%% Main Document %%%%%%%%%%%%%%%%%%%%%%%%%%%%%
\begin{abstract}
Chaperonins are biological nanomachines that help newly translated proteins to fold by rescuing
them from kinetically trapped misfolded states. Protein folding assistance by the chaperonin machinery is obligatory {\it in vivo} for a subset of proteins in the bacterial proteome. Chaperonins are large oligomeric complexes, with unusual seven fold symmetry (group I) or eight/nine fold symmetry (group II), that form double-ring constructs, enclosing a central cavity that serves as the folding chamber. Dramatic large-scale conformational changes, that take place during ATP-driven cycles, allow chaperonins to bind misfolded proteins, encapsulate them into the expanded cavity and release them back into the cellular environment, regardless of whether they are folded or not. The theory associated with the iterative annealing mechanism, which incorporated the conformational free energy landscape description of protein folding, \textit{quantitatively} explains most, if not all, the available data. Misfolded conformations are associated with low energy minima in a rugged energy landscape. Random disruptions of these low energy conformations result in higher free energy, less folded, conformations that can stochastically partition into the native state. Two distinct mechanisms of annealing action have been described. Group I chaperonins (GroEL homologues in eubacteria and endosymbiotic organelles), recognize a large number of misfolded proteins non-specifically and operate through highly coordinated cooperative motions. By contrast, the less well understood group II chaperonins (CCT in Eukarya and thermosome/TF55 in Archaea), assist a selected set of substrate proteins. Sequential conformational changes within a CCT ring are observed, perhaps promoting domain-by-domain substrate folding. Chaperonins are implicated in bacterial infection, autoimmune disease, as well as protein aggregation and degradation diseases. Understanding the chaperonin mechanism and the specific proteins they rescue during the cell cycle is important not only for the fundamental aspect of protein folding in the cellular environment, but also for effective therapeutic strategies.
  

\end{abstract}
\newpage
\section{Introduction}
Protein folding in the cell is not always a spontaneous process due to unproductive pathways of misfolding and aggregation. Chaperonin molecules in bacterium prevent such off-pathway reactions and promote protein folding through spectacular ATP-driven cycles of binding and releasing substrate proteins (SPs). Chaperonins are distinguished among the molecular chaperone family by the presence of a cavity that offers a productive environment for protein folding, thus preventing unwarranted inter protein interactions, which could occur in the crowded cellular environment. Many chaperones are known as heat-shock proteins (Hsp), alluding to their overexpression under stress conditions, although their action is also required under normal cell growth. Availability of chaperone assistance at critical junctures, for example through thermotolerance, ensures cell viability even when cellular functions would otherwise be overwhelmed. More broadly, comprehensive protein quality control relies on a range of chaperone subfamilies, classified according to their molecular weight, to deliver assistance with essential processes in the protein lifecycle: folding/refolding, with an important role for Hsp60/Hsp10 (GroEL/S), Hsp90 and Hsp70/Hsp40 classes (DnaK/DnaJ); protection against oxidative stress, Hsp33; disaggregation, Hsp100 (Hsp104/ClpB), Hsp70/Hsp40, and small Hsps (sHsp); and degradation, Hsp100 (Clp family, p97, the proteasome Rpt1-6 ring) \cite{Parsell1993AnnuRevGenet,Wickner1999Science,Frydman2001AnnuRevBiochem,Kim2013AnnuRevBiochem}.    

\begin{figure}
        \subfigure[]{
        \label{fig:groel}
\includegraphics[width=2.7in,height=3.1in]{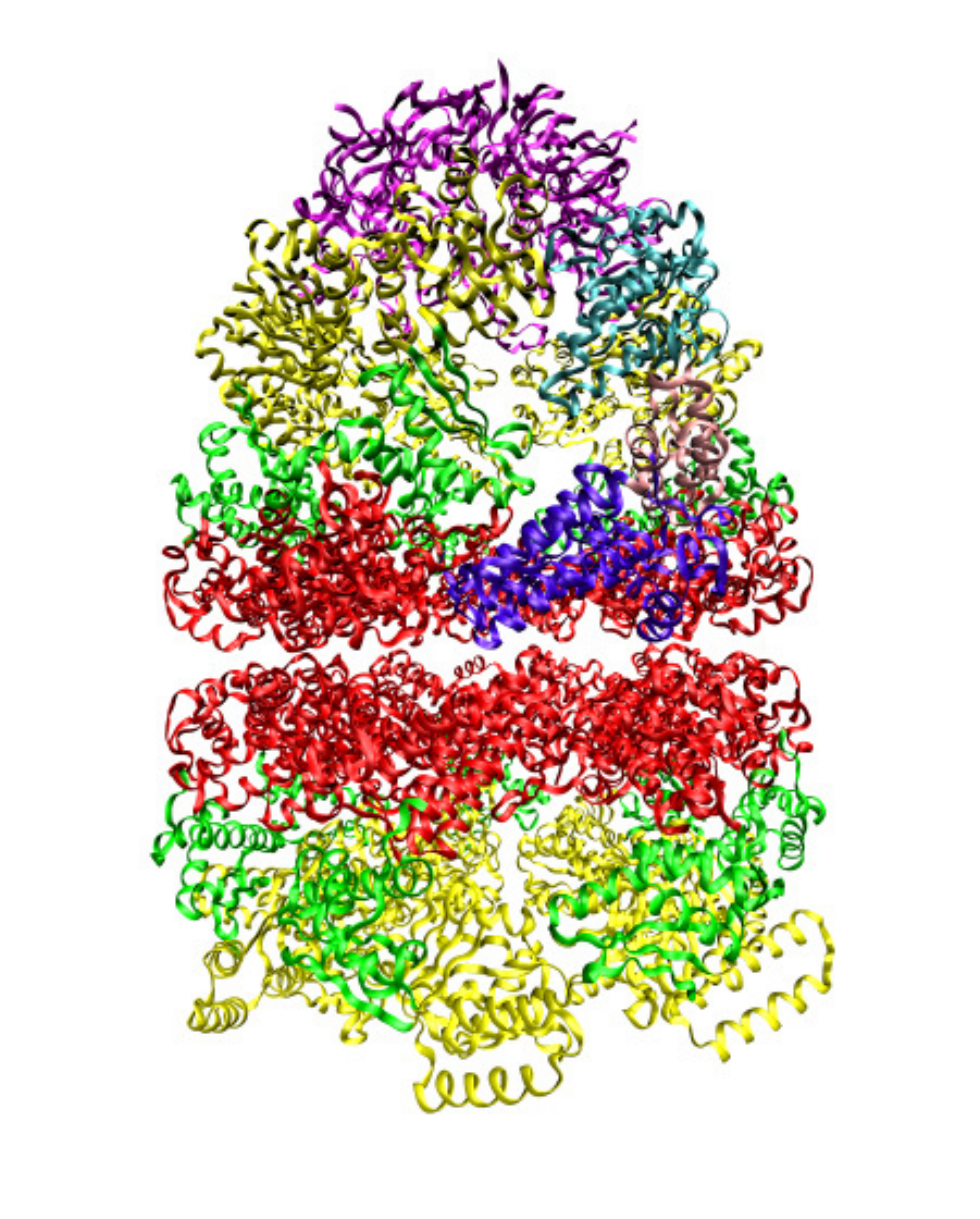}
        }
        \subfigure[]{
        \label{fig:thermosome}
        \includegraphics[width=2.8in,height=2.7in]{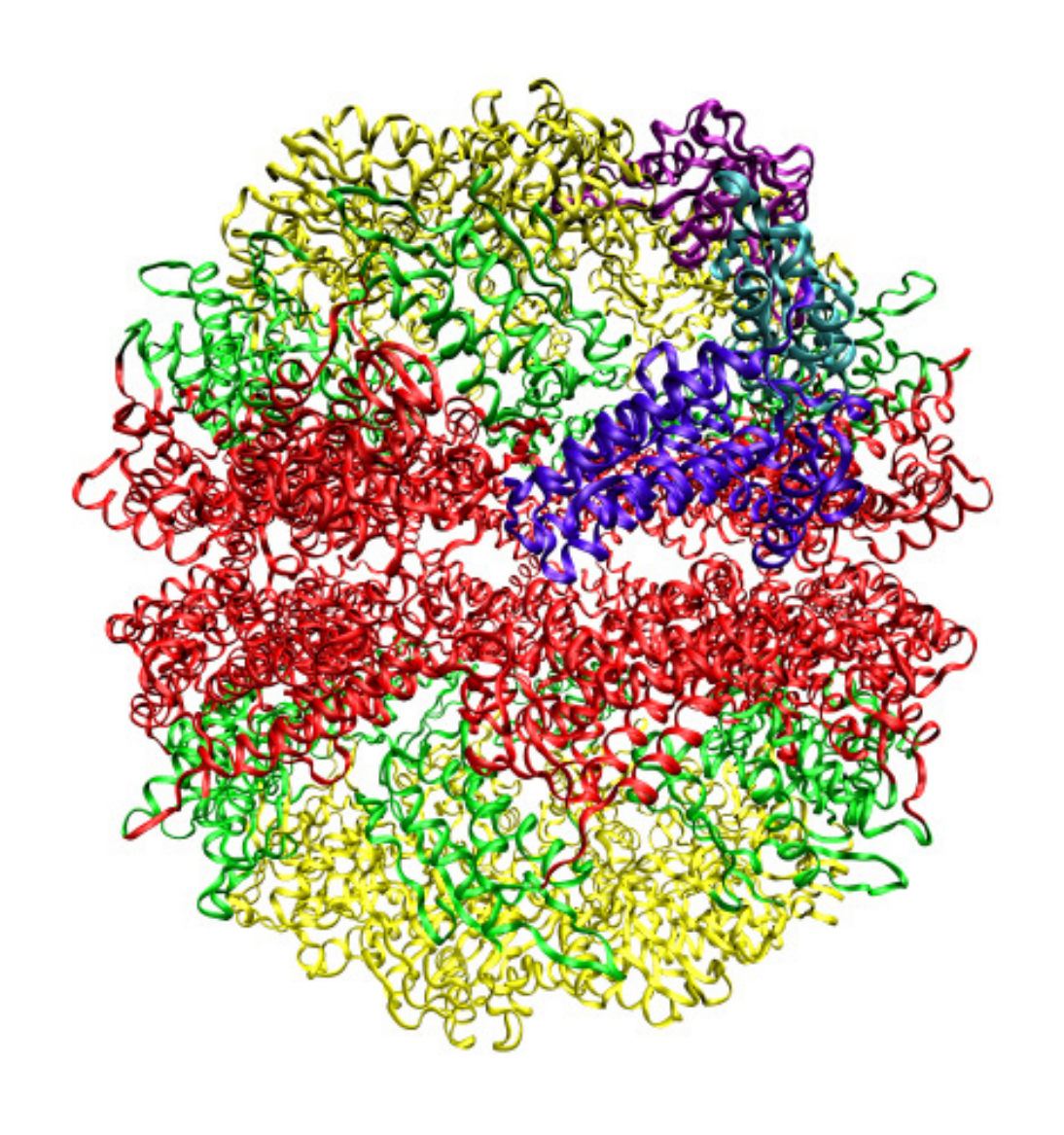}
        }
\caption{\baselineskip 12pt Prototypes of chaperonin classes. (a) Group I chaperonin GroEL and its co-chaperonin GroES (purple) (b) Group II chaperonin thermosome. Three domains, equatorial (red), intermediate (green), and apical (yellow), are distinguished within each subunit. The domains belonging to one subunit of each chaperonin are higlighted. Adapted from Ref. \cite{ditzel98cell}.}
\label{fig:chaperonins}
\vskip -0.1in
\end{figure}

Two distinct chaperonin classes have been identified. GroEL and its co-chaperonin GroES in {\it Escherichia coli} (Figure \ref{fig:groel})  are the prototype for chaperonin systems found in eubacteria and endosymbiotic organelles, or Group I chaperonins. The thermosome (Figure \ref{fig:thermosome}) and TCP-1 ring complex (TRiC, or CCT for chaperonin-containing TCP1) are representative for archaeal and eukaryotic cells, respectively, and are known as Group II chaperonins. Structural characterization \cite{braig94nature,xu97nature,ditzel98cell,Fei2013PNAS} reveals that chaperonins have an oligomeric, double-ring structure, composed of two (thermosome) or more (CCT) distinct subunits within the same ring or identical subunits (GroEL). Within each subunit there are three distinct domains: the ATP-binding equatorial domain, the flexible apical domain and
the intermediate hinge region. The co-chaperonin GroES is a single-ring oligomer with identical subunits, capping one of the GroEL rings (Figure \ref{fig:groel}). This elaborate annealing machinery is present in nearly all organisms and it is essential for cell survival \cite{fayet89jbact}.

Why does folding of some proteins in the crowded cellular milieu require chaperonin assistance? This requirement does not exist {\it in vitro}, as favorable conditions for folding can be identified for known chaperonin substrates. Cellular conditions, however, are unfavorable (non-permissive) for a subset of these proteins, leading to formation of misfolded conformations. To reach the native state from the misfolded conformations, proteins must overcome large free energy barriers, a feat which could prove difficult to accomplish within the biological time scale. Moreover, misfolded proteins expose patches of hydrophobic amino acids, making them potential targets for aggregation or leaving them vulnerable to degradation. Chaperonins rescue proteins trapped in misfolded conformations and allow them to reach the native state within a protected folding chamber.

It should be noted that the chaperone annealing action does not alter the three-dimensional conformation of the native protein, in accord with Anfinsen's hypothesis \cite{anfinsen73science} that the information needed for the native state is encoded solely in the amino acid sequence. Instead, chaperones induce pathways that ensure the correct folding of newly translated or newly translocated proteins \cite{Naqvi2022SciAdv}. 

Here, we provide  our perspective on the substrate recognition mechanisms for the two chaperonin types. A number of reviews describe in detail other fundamental features of the chaperonin machinery, including structure, allosteric motion and ATPase activity \cite{thirumalai01anrevbiop,saibil02tibs,hartl02science,fenton03qrevbiop,spiess04ticb,horovitz05cosb,horwich06chemrev,Gruber2016ChemRev,Horovitz2022AnnuRevBiophys,Thirumalai2020ProtSci}. We refer
the interested reader to these accounts for a broader picture of the chaperonin mechanisms.

We also briefly examine the role of chaperonin in disease and point to extensive research in the area \cite{ranford02molpath}. Prevention of aggregation through chaperonin assisted folding of non-native polypeptides naturally suggests that defects in the chaperonin machinery may result in disease. The extreme situation, the absence of chaperonin, is fatal, as a consequence of the essential nature of this machinery for the cell. Besides these immediate implications, chaperonins are also found to be major immunogens that play an important role in infection, autoimmune disease, and idiopathic diseases such as arthritis and atherosclerosis. Considering the potential therapeutic use, the study of chaperonin assisted protein folding is likely to suggest valuable practical approaches.

\section{Chaperonin hemicycle}

Chaperonins operate as continuous annealing machines by alternating encapsulation of substrate proteins within the cavity of each ring. These encapsulation events are enabled by large scale, coordinated, conformational transitions that take place in conjunction with ATP and GroES binding in the active ring of GroEL. In this section, we focus on the series of events that occur during the GroEL hemicycle.
\begin{figure}
\centering
\includegraphics[width=4.5in,height=3.2in]{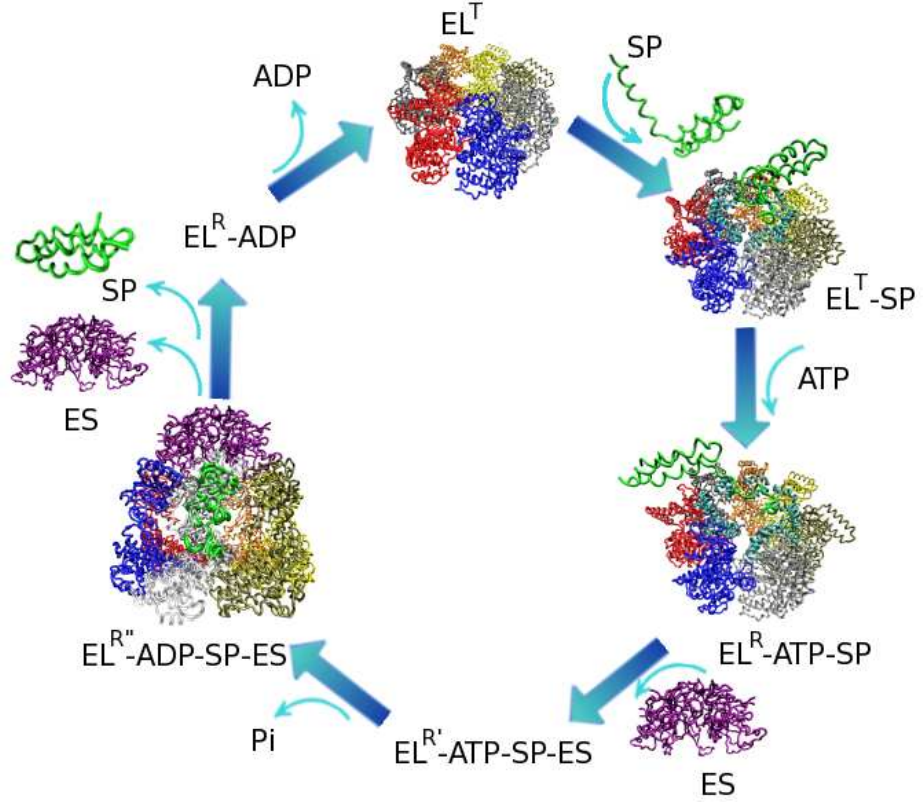}
\caption{\baselineskip 12pt Reaction hemicycle of GroEL illustrating the substrate protein (SP) folding assistance. EL and ES stand for GroEL and GroES respectively. The GroEL T state has a high affinity for SP binding. Upon ATP and GroES binding, the SP is displaced into the expanded GroEL cavity, where productive folding can take place. Dissociation of the complex occurs upon the initiation of a folding reaction in the opposite GroEL ring. The structures of the T, R, and R$^{\prime\prime}$ states are known. Reproduced from Ref. \cite{stan07pnas} \copyright (2007) National Academy of Sciences.}
\label{fig:hemicycle}
\end{figure}
At the initiation of the chaperonin cycle, termed the T state (Figure \ref{fig:hemicycle}), GroEL presents a nearly continuous hydrophobic ring formed at the mouth of the cavity by the seven apical domain binding sites \cite{braig94nature}. This state has high affinity for non-native polypeptides, which also present exposed hydrophobic surfaces. Binding of misfolded proteins to GroEL prevents the formation of irreversible protein aggregates. Upon ATP and GroES binding to the same ring, large-scale, entirely concerted, domain motions in that ring result in doubling the size of the cavity. During these transformations, GroES, which occupies the same apical binding sites as the SP \cite{fenton94nature,xu97nature,buckle97pnas,chen99cell}, displaces the SP in the largely expanded cavity. As a result of these spectacular allosteric transitions, the SP is presented with a completely different, mostly hydrophilic, environment that promotes SP folding. The chaperonin cycle is completed by ATP hydrolysis and the binding of ATP in the opposite ring, which initiates the cycle in that ring. These events trigger the release of GroES, ADP and SP from the folding chamber. Stringent GroEL substrates require several cycles of binding and release in order to reach their native state.  In each cycle, productive folding, if it were to occur at all takes place within the cavity \cite{Thirumalai2020ProtSci}.

\section{Iterative Annealing Mechanism}

\begin{figure}
\centering
\includegraphics[width=5.3in,height=4.in]{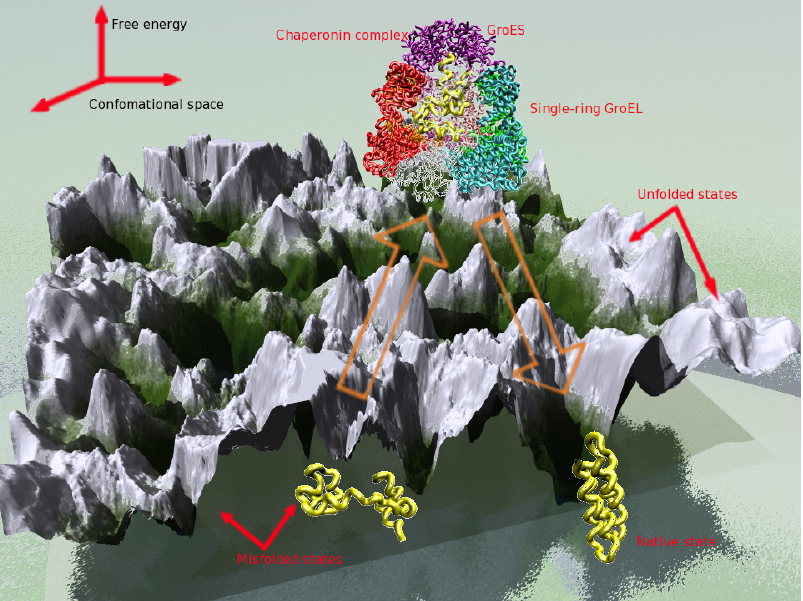}
\caption{\baselineskip 12pt Energy landscape perspective of the chaperonin annealing action. }
\label{fig:landscape}
\end{figure}

The function of the GroEL machinery can be quantitatively understood within the Iterative Annealing
Mechanism (IAM) framework \cite{todd96pnas}. This mechanism is described in the framework of the energy landscape, which associates a free energy to each conformational state of the protein (Figure \ref{fig:landscape}). During each cycle, the SP is rescued from one of the low energy minima, that corresponds to a misfolded state. From the ensuing higher free energy state, the protein chain undergoes kinetic partitioning \cite{guo95biopol} to either the native state or to the same or a different low energy minimum.
Protein folding in a model cavity has been investigated using implicit solvent and coarse-grained models for the SP \cite{betancourt99jmb,klimov02pnas,baumketner03jmb,takagi03pnas,jewett04pnas,vaart04biophysj,stan07pnas}. These studies have provided several important clues about how protein folding occurs in confinement. It turns out that an optimum range of interactions between the cavity wall and the SP results in enhanced stability and folding rates.

\section{GroEL substrate protein binding mechanism}

GroEL manifests a promiscuous behavior towards binding non-native polypeptides. Misfolded proteins, that expose hydrophobic residues,  are recognized by GroEL without preference for a specific secondary or tertiary structure
\cite{viitanen92protsci,aoki00jbc}. Despite the large number of proteins that can form complexes with GroEL \cite{viitanen92protsci}, {\it in vivo} only about 5-10\% of {\it E. coli} proteins can afford to use the chaperonin machinery under normal conditions \cite{lorimer96faseb,ewalt97cell}. Even upon heat stress, only about 30\% of {\it E. coli} proteins require folding assistance \cite{horwich93cell}. The relatively reduced participation of GroEL to protein folding in the cell prompts us to wonder {\it why only a subset of proteins of the entire organism uses chaperonin assistance.} Given the GroEL promiscuity, {\it how does GroEL discriminate between substrates and non-substrates within a proteome?}

Addressing these questions is challenging, from an experimental point of view, because of the inherent difficulty in arresting structures of complexes formed between GroEL and non-native polypeptides. Somewhat surprisingly, even after 25 years, the only available crystal structures of GroEL-bound ligands correspond to the GroEL-GroES complex \cite{xu97nature} and to peptides bound to GroEL
\cite{wang03jmb} or to the GroEL apical domain fragment \cite{buckle97pnas,chen99cell}, while a number of lower resolution cryo-EM structures \cite{roseman01jsb,ranson01cell,falke05jmb,chen06structure} are available. Nevertheless, these structures, as well as a number of biochemical studies, identified the GroEL binding sites and the multivalent binding of stringent substrate proteins. Bioinformatic analyses complementing these studies pinpointed chaperonin signaling pathways and chemical character conservation at functionally relevant sites.

Characterization of the GroEL binding sites using mutational \cite{fenton94nature} and crystallographic \cite{xu97nature} studies pointed towards a mostly hydrophobic groove between two amphiphatic helices (Figure \ref{fig:groes_structure}), as well as a nearby loop, located in the apical domain of each GroEL subunit. Specific residues implicated in GroES and SP binding are Tyr 199, Tyr 203, Phe 204, Leu 234, Leu 237, Leu 259, Val 263, Val 264 \cite{fenton94nature}. In addition, these studies led to the key observation that the same GroEL region responsible for recognizing misfolded substrates is ultimately destined to
form the interface with GroES in the course of the chaperonin cycle. Strikingly, the structures of peptides bound to GroEL overlap significantly \cite{chen99cell}, suggesting that strong restrictions are imposed on the bound conformation.

Bioinformatic analysis of a large number of chaperonin sequences further revealed that the various chaperonin functions (peptide binding, nucleotide binding, GroES and SP release) require that the chemical character and not the identities of specific amino acids be preserved \cite{stan03biopchem}. Moreover, this study lent support to the sequence analysis by Kass and Horovitz \cite{kass02proteins}, which suggested that correlated mutations couple residue doublets or triplets along signaling pathways within GroEL or between GroEL and GroES.

Multivalent binding of stringent SP substrates was suggested to be implicated in the GroEL unfoldase action. This action is brought about by the large scale conformational transitions that take place in coordinated fashion in all GroEL subunits during the chaperonin cycle, resulting in an increased separation of the apical binding sites. At the initiation of the chaperonin cycle, the seven GroEL binding sites form a nearly continuous ring at the cavity opening. Stringent GroEL substrates, such as malate dehydrogenase and Rubisco (not natural substrates for GroEL, however Rubisco is a substrate of the Rubisco binding protein, GroEL homolog in chloroplast), appear to interact with at least three consecutive binding sites \cite{farr00cell,Elad2007MolecularCell}. By contrast, Rhodanese, which is a less-stringent substrate, requires two non-contiguous binding sites \cite{farr00cell}.  The effect of this displacement, corroborated with multivalent SP binding, is to impart a stretching force to the SP
\cite{thirumalai01anrevbiop}.

Taken together, these important results suggest that substrate recognition involves peptides that occupy the GroEL binding site in a similar conformation as the GroES mobile loop. For stringent GroEL SPs, multiple interfaces are formed involving these peptides and several contiguous GroEL subunits. The peptide complementarity to the GroEL binding sites is defined, as in the GroES case, by amino acids whose chemical character is strongly conserved.

\section{Identification of GroEL substrates at the proteome level}

The promiscuous behavior of GroEL towards binding non-native polypeptides appears to be at variance with the relatively small fraction of protein chains in an organism that actually use the GroEL machinery. However, common features of GroEL substrates and similar conformations of bound peptides, as discussed above, suggest a set of requirements for GroEL recognition. Several computational approaches \cite{chaudhuri05csc,stan05protsci,noivirt07bioinfo,Tartaglia2010JMBa,Raineri2010Bioinformatics} and proteomic studies \cite{houry99nature,kerner05cell} were successful in identifying and characterizing GroEL substrates within whole proteome.

Proteomic and biochemical studies \cite{houry99nature,kerner05cell}
provided the first experimental identification of GroEL SPs, on a proteome-wide scale, in {\it E. coli}. These studies found that 252 of the $\sim$2400 cytosolic proteins in {\it E. coli} interact with GroEL. Among this set of proteins, 85 are stringent substrates under normal growth conditions and they occupy 75-80\% of the GroEL capacity. Additional {\it in vivo} studies \cite{chapman06pnas} involving a temperature-sensitive lethal {\it E. coli} mutant suggested a wider set of $\sim$300 GroEL interacting proteins, including some that had not been revealed by previous {\it in vitro} experiments. The latter study raises the possibility that even transient GroEL interaction, in the cellular environment, suffices to prevent aggregation of misfolded proteins. The set of obligate \textit{in vivo} substrates was subsequently narrowed  to $\simeq$ 60 proteins identified in experiments using GroE-depleted conditions \cite{Fujiwara2010EMBOJ}, and completed by the addition of 20 novel substrates identified using cell-free proteomics \cite{Niwa2016FEBSLett}. GroEL substrates were also identified in other bacteria, such as \textit{Thermus thermophilus} \cite{shimamura04struct} and \textit{Bacillus subtilis} \cite{Endo2007BiosciBiotechnolBiochem}.

\begin{figure}
        \subfigure[]{
        \label{fig:groes_structure}
       \includegraphics[width=2.8in,height=2.5in]{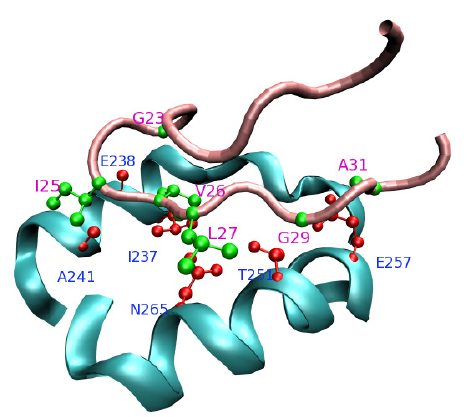}}
        \subfigure[]{
        \label{fig:groes_contact_map}
\includegraphics[width=2.5in,height=2.2in]{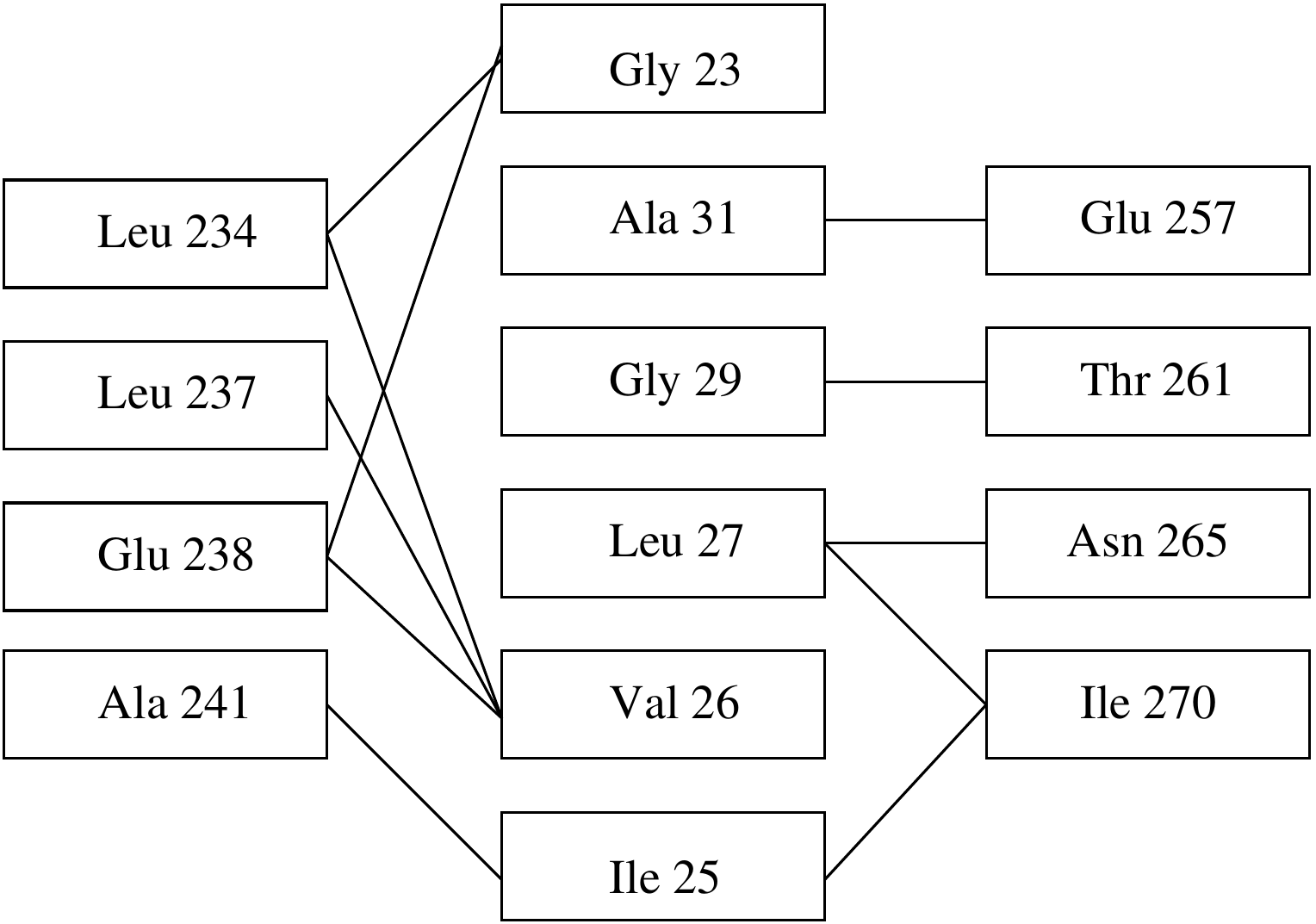}}
\caption{ (a) Contacts between GroEL helices H and I (cyan) and the GroES mobile loop (pink). Sidechains of the residues that form the closest contacts are shown in red (GroEL) and green (GroES). (b) Schematic representation of contacts between GroEL and GroES. Reproduced from Ref. \cite{stan05protsci}.}
\label{fig:contacts}
\end{figure}

One line of computational research focuses on identifying polypeptide regions within proteins that render them natural substrates for GroEL. The underlying hypothesis is that natural SPs have the same sequence complementarity to the GroEL binding site as GroES \cite{chaudhuri05csc,stan05protsci}. Therefore, SPs possess sequence patterns similar to the GroES mobile loop segment 23-31, GGIVLTGAA, which binds to GroEL. In one approach \cite{chaudhuri05csc}, SP binding motifs are defined as strong hydrophobic patches ({\it i.e.} containing amino acids L, V, I, F, M) having 40-50\% sequence similarity to the GroES segment GGIVLTG. The sequence similarity is evaluated using a pairwise alignment between the protein sequence and the peptide GGIVLTG and allowing amino acid substitutions that preserve the chemical character (hydrophobic-hydrophobic or same charge). In a different approach \cite{stan05protsci}, the binding motif is required to match the pattern G\_IVL\_G\_A that includes N$_C$=6 GroES amino acids in contact with GroEL (Figure \ref{fig:contacts}) and three arbitrary amino acids (``\_''). Pattern matching takes into account possible amino acid substitutions that preserve the chemical character, as well as less strongly bound peptides, having four (G\_IVL) and five (G\_IVL\_G) contacts. Natural SPs must possess multiple copies of the binding motif, N$_B$, to satisfy the required multivalent binding to GroEL. About a third of the sequences in the {\it E. coli} proteome are expected to be natural SPs \cite{horwich93cell}. This method retrieves the expected fraction of natural SPs in {\it E. coli} for sequences that satisfy $4<$ N$_C < 6$ and 2 $<$ N$_B < 4$. No preferred secondary structure emerges in this set of proteins. This method is able to identify 80\% of experimentally determined natural substrate proteins for GroEL from {\it E. coli} \cite{houry99nature,kerner05cell} and predicted SPs in several other proteomes.

GroEL must not only recognize proteins that require folding assistance, but also the protein conformations that must be remodeled. How does GroEL discriminate between native conformations, which it should not recruit, from the misfolded conformations of proteins it must selectively assist? A structural and bioinformatic analysis \cite{stan06pnas} found that GroES-type binding motifs are not significantly exposed to solvent in the native conformation of GroEL SPs. This result suggests that GroEL recognition of misfolded conformations of SPs requires that multiple GroES-type binding motifs be solvent-exposed. In accord with this hypothesis, molecular dynamics simulations that probe extensively the conformational space of an obligate GroEL substrate, DapA \cite{Nagpal2015PLoSComputBiol}, reveal that its GroES-type binding motifs are solvent-exposed in unfolding intermediates, but are inaccessible in the native conformation. These studies find that, for the seven motifs identified within the DapA sequence, the average solvent-exposed area per residue increases from $\simeq$ 74\AA\ in the native conformation to $\simeq$ 182\AA\ in the intermediate structures. Experimental studies using hydrogen-exchange coupled with mass spectrometry \cite{Georgescauld2014Cell} support the increased exposure of the hydrophobic segments and loss of hydrogen bonds that accompany the destabilization of the TIM-barrel core of this substrate. 

A different line of computational research \cite{noivirt07bioinfo,Azia2012FEBSJ,Raineri2010Bioinformatics,Tartaglia2010JMBa} uses machine learning approaches to examine physicochemical characteristics of {\it E. coli} proteins that indicate a requirement for GroE-dependent folding. Among two sets of {\it in vivo} substrates \cite{kerner05cell,chapman06pnas}, stringent dependence on GroEL correlates with low folding propensity and high translation efficiency \cite{noivirt07bioinfo}. Secondary structure content, as well as contact order, which quantifies the average distance along the polypeptide chain between amino acids that form native contacts, were not found to distinguish GroEL SPs from other proteins. Consistently, this study found that homologues of these SPs in {\it Ureaplasma urealyticum}, an organism that lacks a chaperonin system, do not possess sequence characteristics that would require them to recruit the GroE system. Additional features were found by two other studies to separate GroEL SPs from GroE-independent proteins. One found that lower rate of evolution, hydrophobicity, and aggregation propensity are characteristics of GroEL SPs \cite{Raineri2010Bioinformatics}, however it was later argued that the estimation of aggregation propensity may reflect the algorithm bias towards amyloid structure \cite{Azia2012FEBSJ}. Solubilities of \textit{E. coli} proteins are found to display a bimodal distribution within a cell-free system in the absence of chaperones, with stringent GroEL SPs belonging to the more aggregation-prone set \cite{Niwa2009PNAS}. In agreement with these results, the second computational approach was successful in distinguishing the GroEL requirement for the previously identified substrate classes \cite{kerner05cell} on the basis of decreasing folding propensity and increasing likelihood of aggregation \cite{Tartaglia2010JMBa}. To probe the substrate requirements in controlled fashion, recent experiments used computationally designed substrates based on the enhanced green fluorescence protein (eGFP) \cite{Bandyopadhyay2017JBC,Bandyopadhyay2019BiophysJ}. These \textit{in vitro} and \textit{in vivo} studies showed that GroEL dependence of eGFP variants increases with increasing frustration \cite{Ferreiro2018COSB}, effected through point mutations\cite{Bandyopadhyay2017JBC}, or contact order \cite{Plaxco1998JMB}, engineered through circular permutations\cite{Bandyopadhyay2019BiophysJ}. Intriguingly, as noted above, \textit{in vivo} GroEL SPs are not distinguishable from non-substrates through the contact order parameter. This suggests that other features play a larger role in determining GroE-dependence.

\section{Specific recognition of substrate proteins by group II chaperonins}

In contrast to the extensive knowledge of the set of proteins that require assistance from the GroEL-GroES system, relatively little is currently known about the substrates of group II chaperonins. The presence of distinct subunit types within group II chaperonins suggests that specialized binding mechanisms were developed to target different substrates. However, the extent of subunit heterogeneity varies among members of this class. In archaeal chaperonins, one \cite{knapp94jmb}, two \cite{waldmann95ejb}, or three \cite{archibald02jmb} distinct subunit types are identified, whereas in eukaryotic chaperonins eight different subunits are described \cite{liou97emboj}. Correspondingly, it is plausible that different substrate recognition mechanisms are used by archaeal and eukaryotic chaperonins. Group II chaperonins have been suggested to employ a sequential, rather than cooperative, mechanism for conformational transitions, consistent with their suggested domain-by-domain folding of SPs and specific SP interaction.

Archaeal chaperonins are abundant in the cell (approximately 1-2\% of cellular proteins) and have low subunit heterogeneity as a result of gene interconversion \cite{archibald02jmb}. These facts prompted the suggestion that, like GroEL, they assist folding of a large set of proteins perhaps through a promiscuous mechanism. In support of this hypothesis, it is noted that thermosome, which has two subunit types, assists folding of GroEL substrates green fluorescence protein \cite{yoshida02jmb} and cythrate synthase \cite{iizuka04jbc}. The coexistence of group I and group II chaperonins within the archaebacterium {\it Methanosarcina mazei} \cite{klunker03jbc} provides an unique opportunity to compare and contrast the annealing action of the two chaperonin classes. Both chaperonins contribute to the folding of ~13\% of the proteins in the archaeal cytosol, albeit the two sets of substrates are non-overlapping \cite{Hirtreiter2009MolMicrobiol}.

The less abundant eukaryotic chaperonin CCT (0.1\% of cellular proteins) uses a significantly different mechanism of substrate recognition than GroEL. CCT was initially suggested to interact only with actins and tubulins \cite{kubota94currbiol}. Recently, numerous other substrates have been identified, including some that contain tryptophan-aspartic acid repeats \cite{spiess04ticb}. Substrates include the myosin heavy chain, the Von Hippel-Lindau (VHL) tumor suppressor, cyclin E, and the cell division control protein. Charged residues on the surface of CCT SPs appear to be required for recognition by the eukaryotic machinery. Intriguingly, CCT substrates cannot be folded by other prokaryotic or eukaryotic chaperones
\cite{tian95nature}.

A challenging aspect of the CCT substrate recognition mechanism is the lack of knowledge of the CCT binding site. Several proposals exist regarding the localization of CCT binding sites. One assumes structural homology to the GroEL binding sites, formed by two apical domain helices \cite{xu97nature}. In contrast to  the GroEL binding site, the two CCT helices have a mostly hydrophilic character, which would be consistent with the notion that CCT recognizes surface charged residues \cite{Jayasinghe2010Proteins}. A second proposed CCT binding site involves a flexible helical protrusion \cite{Heller2004JMB} that acts as a built-in lid for the chaperonin cavity. Finally, the inner side of the closed cavity was also suggested as CCT binding site \cite{pappenberger02jmb}. This region has a mostly charged and polar character, a feature similar to the lining of the GroEL cavity wall. At this time, few experimental data are available to unambiguously define the CCT binding site. A study that used photocrosslinking and fluorescence spectroscopy to probe VHL binding \cite{spiess06molcell}, provides strong indication that the CCT binding sites are located within helix 11, which is structurally homologous to the GroEL binding site.

It is possible that more than one of these proposed locations correspond to {\it in vivo} CCT binding sites. This would not be
completely surprising given the diversity among CCT substrates and the CCT inhomogeneous oligomeric structure \cite{spiess04ticb}. Distinct CCT subunits may serve the purpose of providing the versatility to recognize different substrates.

\section{Implication of chaperonins in disease}

An intriguing connection was made between the Hsp60 chaperonin class and prion disease \cite{debburman97pnas}. Prion proteins are suggested to form fibrillar aggregates upon conversion of the normal cellular form PrP$^C$, having primarily $\alpha$-helical structure, into a $\beta$-sheet rich misfolded conformation, PrP$^{Sc}$. Experiments performed {\it in vitro}
found that GroEL promotes conversion to the disease-related PrP$^{Sc}$ \cite{debburman97pnas}. These authors proposed that {\it in vivo} validation of the chaperonin-assisted conversion would provide a natural target for clinical approaches.

Mutations in human chaperonins result in diseases, such as the hereditary spastic paraplegia \cite{hansen02amjhg}, or the McKusick-Kaufman Syndrome \cite{stone00natgen}. Chaperonins have been implicated, through autoimmune response, as putative causes of diseases such as rheumatoid arthritis, atherosclerosis, and inflammation \cite{ranford00exprevmm}. Immunosuppresive action of chaperonins has been described in animal models of juvenile arthritis \cite{eden89rheumint} and diabetes \cite{elias90pnas}, as well as in human pregnancy \cite{cavanagh94ejb}. Immunization with a mycobacterial chaperonin was suggested to protect against arthritis \cite{eden91immrev}. To date, there is no clear understanding of the subset of chaperonin SPs affected by these mutations and the precise effect of these mutations on chaperonin annealing action \cite{barral04scdb}. Mastering the intricacies of the chaperonin action will provide answers to these questions and suggest effective therapies.

\section{Conclusions}

Protein folding assistance mediated by chaperonins is a critical quality control mechanism to maintain protein homeostasis. Selective recruitment of substrate proteins by chaperonins represents a fundamental regulatory step in the remodeling action, given the limited availability of chaperonins within the cytosol and the stringent dependence of a subset of proteins on this assistance. Remarkably, GroEL substrate selectivity is achieved even as the chaperonin promiscuously binds misfolded proteins. As highlighted here, research efforts to elucidate the substrate recognition mechanism have primarily focused on two complementary questions. One question is focused on how GroEL binds substrate proteins that require its assistance. As the GroEL binding site is well established and the GroES co-chaperone competes with substrates during the chaperonin cycle, this suggests that natural SPs include polypeptide regions similar to the GroES loops that participate in the interface with GroEL. The additional observation that substrates interact with multiple GroEL subunits (2-3) further defines the requirement that several GroES-type motifs be present within the polypeptide chain, at least for stringent substrates. The other question refers to which proteins are likely to require folding assistance \textit{in vivo}. Here, low partition factor (fraction of molecules that fold spontaneously) and high aggregation propensity emerge as important factors that underlie the GroEL requirement. In addition, such factors can help to explain the extent of GroEL dependence among known substrates.

\section{Acknowledgments}

The authors gratefully acknowledge stimulating comments from Amnon Horovitz. 

%This work was supported by the National Science Foundation (CHE 19-00033) and the Welch Foundation (F-0019) through the Collie-Welch Regents Chair.

\section*{Conflict of Interest Statement}
%All financial, commercial or other relationships that might be perceived by the academic community as representing a potential conflict of interest must be disclosed. If no such relationship exists, authors will be asked to confirm the following statement: 

The authors declare that the research was conducted in the absence of any commercial or financial relationships that could be construed as a potential conflict of interest.

\section*{Author Contributions}

All authors conceived and executed the project, and wrote the report.
%The Author Contributions section is mandatory for all articles, including articles by sole authors. If an appropriate statement is not provided on submission, a standard one will be inserted during the production process. The Author Contributions statement must describe the contributions of individual authors referred to by their initials and, in doing so, all authors agree to be accountable for the content of the work. Please see  \href{https://www.frontiersin.org/about/policies-and-publication-ethics#AuthorshipAuthorResponsibilities}{here} for full authorship criteria.

\section*{Funding}

DT is grateful to the National Science Foundation (CHE 19-00033) for support. Additional support from
the Welch Foundation (F-0019) through the Collie-Welch Regents Chair is greatly acknowledged. GS would like to acknowledge the National Science Foundation (MCB-2136816) for support.
%Details of all funding sources should be provided, including grant numbers if applicable. Please ensure to add all necessary funding information, as after publication this is no longer possible.

\providecommand{\latin}[1]{#1}
\makeatletter
\providecommand{\doi}
  {\begingroup\let\do\@makeother\dospecials
  \catcode`\{=1 \catcode`\}=2 \doi@aux}
\providecommand{\doi@aux}[1]{\endgroup\texttt{#1}}
\makeatother
\providecommand*\mcitethebibliography{\thebibliography}
\csname @ifundefined\endcsname{endmcitethebibliography}
  {\let\endmcitethebibliography\endthebibliography}{}

\end{document}